# Quantitative Phase Retrieval and Characterization of Magnetic Nanostructures via Lorentz (Scanning) Transmission Electron Microscopy

Kayna L. Mendoza[1,2], Haoyang Ni[3], Georgios Varnavides[4,5], Miaofang Chi[6], Colin Ophus[5], Amanda Petford-Long[1,2], Charudatta Phatak[*2]

[1]Department of Materials Science and Engineering, Northwestern University, Evanston, IL, USA.
[2]Materials Science Division, Argonne National Laboratory, Lemont, IL, USA.
[3]Department of Materials Science and Engineering, University of Illinois at Urbana-Champaign, Urbana, IL, USA.
[4]Miller Institute for Basic Research in Science, University of California, Berkeley, CA, USA.
[5]National Center for Electron Microscopy, Lawrence Berkeley National Laboratory, Berkeley, CA, USA.
[6]Center for Nanophase Materials Sciences, Oak Ridge National Laboratory, Oak Ridge, TN, USA.

E-mail: `cd@anl.org`

**Abstract.** Magnetic materials phase reconstruction from Lorentz transmission electron microscopy (LTEM) measurements has traditionally been achieved using longstanding methods such as off-axis holography (OAH) and the transport-of-intensity equation (TIE). Amidst the increase in access to processing power and the development of advanced algorithms, phase retrieval of nanoscale magnetic materials with higher fidelity and resolution, potentially down to the few nanometer limit, becomes possible. Specifically, reverse-mode automatic differentiation (RMAD) and the extended electron ptychography iterative engine (ePIE) are two methods that have been utilized for high confidence phase reconstructions using LTEM through-focal series imaging and Lorentz scanning TEM (Ltz-4D-STEM), respectively. This work evaluates phase retrieval using TIE, RMAD, and ePIE in simulations consisting of an array of Permalloy ($Ni_{80}Fe_{20}$) nanoscale islands. Extending beyond simulations, we demonstrate total phase reconstructions of a NiFe nanowire using OAH and RMAD in LTEM and ePIE in Ltz-4D-STEM experiments and determine the magnetization saturation through corroborations with micromagnetic simulations. Finally, we show how the total phase shift gradient can be utilized to observe and characterize the proximity effects emanating from neighboring magnetic island interactions and an isolated NiFe nanowire.

# 1 Introduction

Targeted explorations of geometrically confined magnetism in nanostructures is of fundamental importance and of considerable interest in the fields of materials science and engineering, condensed matter physics, chemistry, as well as biology owing to magnetic characteristics that exhibit complex properties and topological texturing. Spin frustration, proximity effects, and novel domain behaviour occurs across length scales from three-, two-, and one-dimensional (3D, 2D, 1D) structures, such that the magnetic characteristics become more prominent at the few hundreds of nanometers range down to a few nanometers [1–4]. In addition to fundamental interest, a critical understanding of the behavior of magnetic nanostructures at these length scales is necessary for engineering novel materials systems for the next generation of energy-efficient computing and storage technologies including spintronic devices [5–7], quantum computing [4, 8, 9], and high-density information storage [9–11].

One approach to understanding the behavior of nanoscale magnetic materials is to observe their magnetic structure with Lorentz transmission electron microscopy (LTEM) [12, 13]. Coupled nanoscale magnets such as artificial spin ices exhibit an expansive range of spin frustration and proximity effects as a function of separation, angle, neighbors, and geometry such that the magnetic susceptibility and ordering are, or can be, inherently affected to minimize the overall energy of the system [1, 14, 15]. Interestingly, domain behaviour evolves into novel territory as dimensions are reduced beyond the few hundreds of nanometers down to the tens of nanometer range. These nanoscale structures can exhibit formation of double vortices, single domains, uniform magnetization, or double helices as a function of geometry and neighboring magnets, as is the case for shape-patterned island arrays and nanowires [16, 17]. Magnetization reversal, domain pinning and/or nucleation, alongside crystallographic and magnetic ordering are behaviours that can be observed with relatively high spatial resolution using LTEM and made quantitative utilizing advanced phase retrieval techniques alongside correlations with micromagnetic simulations. High spatial resolution and high phase sensitivity is imperative for quantization of the total phase and subsequent extraction of the magnetic contributions in order to better understand the energy mechanisms that control the complex magnetic properties. As such, LTEM enables high-resolution characterization of the magnetic structure in a material at sub-10 nm length scales [12, 13, 18]. This requires operating in a magnetic field-free region in the LTEM instrument, which can either be achieved by utilizing the field free miniature lens or, more recently, a dedicated field-free objective lens. Thus, magnetic contrast arises from the interaction of the transmitted electron beam through the object's electromagnetic potentials and results in a shift of the total phase ($\phi_t$) of the electron wavefunction which is described by the Aharanov-Bohm relation in Equation 1 [19].

$$\begin{aligned}\phi_t(\boldsymbol{r}_\perp) &= \phi_e(\boldsymbol{r}_\perp) + \phi_m(\boldsymbol{r}_\perp); \\ &= \frac{\pi}{\lambda E}\int V(\boldsymbol{r}_\perp)d\mathbf{l} - \frac{\pi}{\phi_0}\int A(\boldsymbol{r}_\perp)\cdot d\mathbf{l}\end{aligned} \quad (1)$$

Here, the contributions to the electromagnetic potential consist of the electrostatic scalar potential $V(\boldsymbol{r})$ and the magnetic vector potential $\mathbf{A}(\mathbf{r})$. $\lambda$ is the electron wavelength, $E$ the relativistic electron energy, $\mathbf{r}_\perp$ is the two-dimensional position vector in the projected plane perpendicular to the direction of electron propagation $\mathbf{l}$, and $\phi_0 = \frac{h}{2e}$ is the magnetic flux quantum. Together, the electrostatic $\phi_e$ and magnetic $\phi_m$ components contribute to the total phase shift $\phi_t$. The individual contributions can be separated by acquiring two sets of measurements where the sign of the magnetic contribution is reversed – most simply achieved by flipping the sample upside down – and adding/subtracting the two measurements to obtain estimates for the electrostatic and magnetic contributions, respectively. This methodology provides the foundation for phase retrieval in LTEM via traditional methods of off-axis holography (OAH) and through-focal-series imaging, as well as newer methods involving reciprocal-space imaging.

Off-axis holography is a longstanding and reliable method for phase retrieval utilizing LTEM [20–22], but requires the use of an electron biprism to directly apply an electrical potential to create an interference pattern. A challenge of OAH is that it requires a region of 'free space' adjacent to the sample area and is therefore not appropriate for extended samples. When these conditions are met, OAH method yields high spatial resolution down to 3.5 nm and moderate sensitivity to magnetic induction on three-dimensional (3D) nanowires [17, 23].

Alternatively, one can utilize the Fresnel contrast which arises during through-focal-series imaging to directly relate magnetic features in the sample, i.e. domain walls. The transport-of-intensity (TIE) equation and reverse-mode automatic differentiation (RMAD) are two methods of phase retrieval that utilize Fresnel contrast in a through-focus dataset. TIE is more straightforward, in that, the phase can be retrieved directly by solving the partial differential equation (PDE) of the image stack. This method is significantly dependent upon alignment of the images in the stack; thus, requires establishing appropriate boundary conditions [12, 24]. An underlying assumption in the TIE formalism is the linearity of the microscope transfer function, which changes with defocus and spatial resolution. This approach is exact at small defocus lengths and while it can be applied at higher defocus values, it severely impacts the achievable spatial resolution [13, 18].

Recent development of an advanced algorithm based on iterative RMAD for phase retrieval shows significant improvements to the achievable spatial resolution, even at high defocus conditions [25]. While RMAD has been reported to retrieve the phase with high spatial resolution and phase accuracy in micron-scale magnetic islands, reaching beyond these size limits to the nanoscale has not been thoroughly reported.

Four-dimensional scanning transmission electron microscopy (4D-STEM) is a widespread technique whereby an image of the diffracted probe intensity is recorded at each scan position. While the diffraction intensities themselves are phase-less, the redundant scattering information encoded in overlapping regions of the diffracted intensities can be retrieved using various 4D-STEM phase-retrieval techniques [26]. The Aharonov-Bohm relation extends to 4D-STEM when operated in Lorentz-mode (Ltz), herein referred to as *Ltz-4D-STEM*, wherein the instrument is operated in low magnification mode [27].

One increasingly common phase retrieval technique applied to Ltz-4D-STEM data is to use electron ptychography, where the far-field diffraction patterns provide information on the spatial frequencies that are locally present in and around the object. Development of advanced algorithms based on the iterative ptychographic engine (ePIE) [28, 29] or on stochastic gradient-descent have reported near-atomic and atomic scale resolution of total phase retrieval with high spatial resolution and phase accuracy [30–33]. Morevover, the technique has recently been applied to magnetic phase shift retrieval in an experimental iron-germanium (FeGe) nanoflake and simulated antiferromagnetic nickel-oxide (NiO) lattice [34, 35].

In this work, we demonstrate retrieval of the total phase shift from simulated images of Permalloy ($Ni_{80}Fe_{20}$) nanoscale islands, calculated from micromagnetic simulations. We focus on the phase retrieval techniques mentioned above, using the TIE formalism via the PyLorentz open-source software [18], RMAD via automatic differentiation LTEM (ADLTEM) open-source python code [25], and ePIE [29]. We then demonstrate how phase retrieval can be applied to experimental data utilizing OAH and RMAD in LTEM and electron ptychography using py4DSTEM [32, 36] in Ltz-4D-STEM. Ultimately, we highlight a range of conditions for the aforementioned phase retrieval techniques using both LTEM and Ltz-4D-STEM instruments, for which each is optimal while also noting their limitations.

## 2 Results

### 2.1 Nanomagnetic Simulation & Phase Retrieval

Micromagnetic simulations of Permalloy nano-scale island arrays were carried out, using the open-source software MuMax [37], to generate the data from which to perform phase retrieval using TIE, RMAD, and ePIE. The simulation data, phase retrieval, and line plots are depicted in Figure 1. The magnetization configuration of the smallest features of the Permalloy nano-island array with dimensions of $220 \times 86 \times 20$ nm ($l \times w \times h$) and 56 nm spacing between adjacent islands, shown in Fig. 1(a), is used as the input vector data for ground truth phase generation via PyLorentz' Mansuripur algorithm which is highlighted in Fig. 1(b). Recall, the total phase shift arises from contributions containing the electrostatic scalar potential and the magnetic vector potential. Thus, mapping the gradient of the total phase, as shown for the ground truth in Fig. 1(c), provides depiction of the total projected magnetic induction. This also provides a measure of the spatial resolution based on the edges of the nanoislands where there is a sharp change in the electrostatic potential, which is evidenced in the line plot of Fig. 1(d). Subsequently, a simulated LTEM through-focal-series image stack was generated using PyLorentz at a defocus condition of ±100 $\mu$m. Phase retrieval was performed on this dataset using TIE and RMAD independently to highlight the standalone results of using a large defocus value, which is evidenced in Fig. 1 (e, f) and (g,h) as the total phase shift and the phase gradient, respectively.

Electron ptychography, on the other hand, for magnetic imaging utilizes a single 4D-dataset taken at a defocus condition in Lorentz-mode that ensures sufficient probe overlap while also maintaining relatively high resolution in real-space. The total phase shift and phase gradient are presented in Fig. 1 (i, j), respectively.

A quantitative assessment of the reconstructed phases was performed by calculating a structural similarity index (SSIM) measurement using the open-source python scikit-image package, SSIM function [38]. The structural similarity index refers to the attributes in an image that represent the structure of the objects that are independent of luminance and contrast, in this case referring to the magnetic islands. The SSIM percentages are depicted for both the total phase shift and phase gradient as shown in Fig. 1 (e-j). Additionally, the proximity effect is evident in the phase gradient via the faint contrast signals emanating from the ends of the magnetic nanostructures while noting that the signals vary in strength. TIE, in comparison with RMAD and ePIE, denotes a lower SSIM index for the reconstructed total phase shift and phase gradient measuring 88.60 % and 93.18 %, respectively. It is apparent, as evidenced by blurring in Fig. 1(e), that the low spatial frequencies are not effectively accounted for in the phase shift reconstructions, unlike RMAD and ePIE. The electrostatic phase appears to be a prominent factor in the TIE reconstruction at a large defocus value that becomes more evident in the phase gradient, as can be seen by the spread of the low spatial frequencies along the nanostructure edges, Fig. 1(f). This is further evidenced by the line plot depicted in Fig. 1(d), such that the TIE curve yields a broadening of the peak indicative of low signal intensity and spatial resolution. Interestingly, the weak phase variations that appear as 'stray fields' in TIE are in relatively significant agreement with the ground truth measuring 99.48 % and 92.34 % at Fig. 1(f) position 1 and 2, respectively. RMAD yields the highest SSIM index to the ground truth, approximately 94.87 % for the total phase shift and even greater at 98.995 % for the phase gradient when assessing the entire field-of-view. Indeed, the RMAD curve plotted in Fig. 1(d) displays the reconstructed phase gradient with the highest spatial frequency and signal intensity. However, the ptychography reconstruction using ePIE displays a near-exact trend, although with a slight shift in the peaks to the left. Yet, in assessing the weak phase variations at position 1 and 2 shown in Fig. 1(h), there is significantly greater variation from the ground truth which measures 79.84 % and 67.74 %, respectively. Contrary to TIE and RMAD, ePIE exhibits the greatest consistency among the quantitative analyses with the ground truth for the total phase shift and phase gradient, yielding an SSIM of 93.36 % and 98.24 %, respectively. The weak phase variations as highlighted by Fig. 1(j) positions 1 and 2 yield a SSIM index of 98.06 % and 95.97 %, respectively.

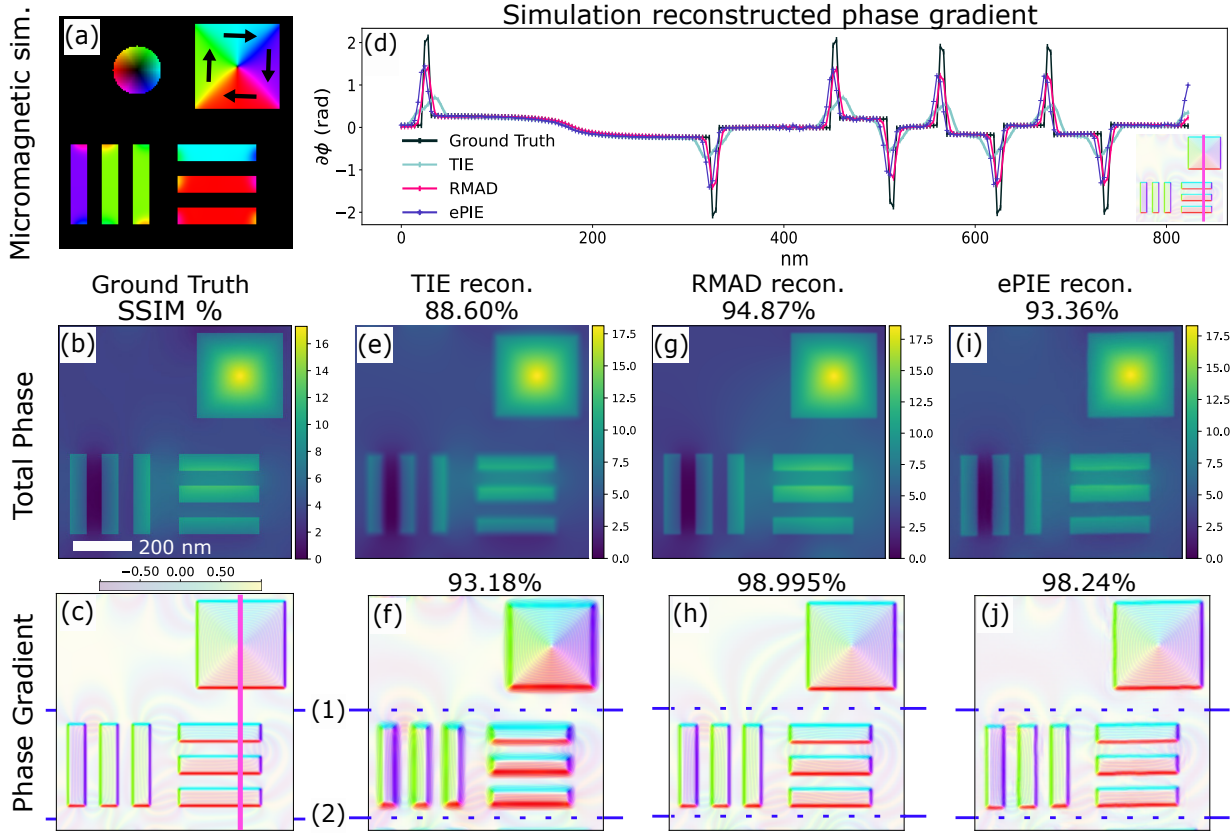

Figure 1: **Comparative performance of phase retrieval techniques on simulated data.** (a) Simulated magnetization distribution in Permalloy nanoscale islands used for phase reconstructions. The total phase shift (b) and the phase gradient (c) are generated using the Mansuripur algorithm. Simulated phase reconstructions depict the total phase shift using TIE (d), RMAD (e), and ePIE (f) alongside the respective phase gradients using TIE (g), RMAD (h), and ePIE (i). The phase gradient is plotted (j) from the line extracted in (b). Scalebar shown in (b) is 200 nm and applicable to all images.

Furthermore, we have extracted and analyzed line plots, as shown in Figure 2, on both axes, x and y, since sensitivity to minor fluctuations can be present. Plot (1) of Fig. 2 presents reconstructions of TIE, RMAD, and ePIE that trend to the ground truth well. However, it is most evident in the TIE reconstruction that there is a loss in spatial accuracy as indicated by the broadening of the gradient in each phase (step) change whereas RMAD and ePIE yield a greater sensitivity to the phase. In plot (2) of Fig. 2, a similar trend of peak broadening within TIE is evident throughout the plot. RMAD, contrary to plot (1), yields a variation in the phase of approximately 0.75 to 1.25 radians. This occurs within the range of 450 to 700 nm profile length and does not remain consistent with the ground truth. As for ePIE, the total phase in both line plots appear to trend well with the ground truth in both phase accuracy and spatial resolution.

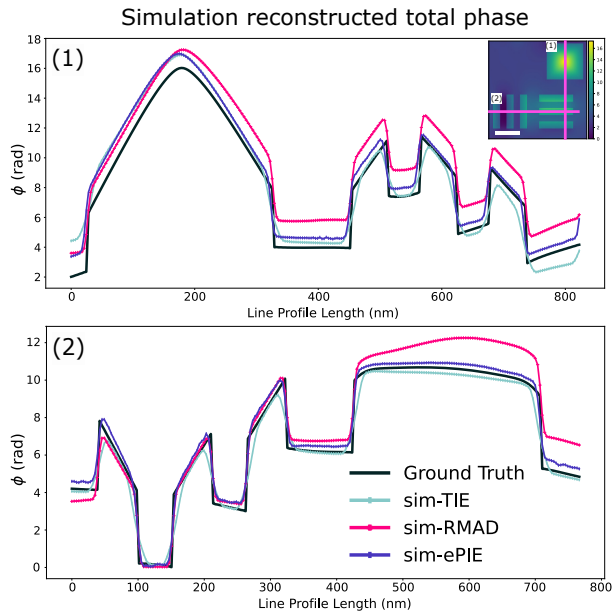

Figure 2: **Simulated total phase reconstruction** of the vertical line (1) from top to bottom and the horizontal line (2) from left to right. Scalebar within inset is 200 nm.

## 2.2 Experimental Phase Retrieval

In order to explore experimental phase retrieval using these different approaches, we imaged a NiFe nanowire that is 36 nm in diameter. First, we collected LTEM measurements to explore phase retrieval via OAH and RMAD. Then, we collected Ltz-4D-STEM measurements to explore phase retrieval via electron ptychography and tilt-corrected bright field STEM using py4DSTEM. Within the various STEM phase-retrieval methods offered by py4DSTEM, we explored we explored iterative ptychography using stochastic gradient descent (of which ePIE is a limiting case) and tilt-corrected BF-STEM (or parallax imaging) [32, 36]. We then compare our experimental data with simulations of a NiFe nanowire carried out using MuMax [37]. The magnetization configuration in a cylindrical Permalloy nanowire of diameter 36 nm and length 105 times greater was initialized along the length of the nanowire. Similar to the previous simulated data in Section 2.1, the ground truth phase was calculated using PyLorentz' Mansuripur-based algorithm.

### 2.2.1 Lorentz Transmission Electron Microscopy

The experimental parameters are presented in Section 5. Figure 3 illustrates the use of LTEM to perform electron OAH and through-focal-series imaging to retrieve phase images of the NiFe nanowire. Bright-field in-focus and defocused images, and the corresponding electron hologram, are shown in Fig. 3 (a-c), respectively. Dunin-Borkowski et al. reported that charging effects can arise at the condenser aperture plane and consequently, have an impact on the total electron phase shift [39]. Thus, a reference hologram was acquired under the same imaging conditions to enable removal of these contributions via background subtraction. The experimental holograms were analyzed utilizing the electron holography functionality in the open-source python package Hyperspy [40]. The field-of-view in the through-focal-series dataset is constrained to near one end of the nanowire as a result of carbon build-up that occurred during our OAH and electron ptychography experiments. A series of 21 defocus conditions were used, including 1 in-focus image and 10 under- and over-focus values ($\Delta f = \pm$ 2, 6, 12, 22, 25, 41, 46, 69, 92, and 134 $\mu$m). Investigations were carried out by varying the quantity of images and selection of defocus conditions utilized in a reconstruction of 200,000 iterations. We found that RMAD reconstructions utilizing a mixed defocus series consisting of two images at moderate (11.52 and 21.89 $\mu$m) and lower (2.30 and 5.76 $\mu$m) defocus values improved the phase retrieval significantly, as shown in Supplementary Figure 2 (c) and is in agreement with reports by Zhou et al. [25].

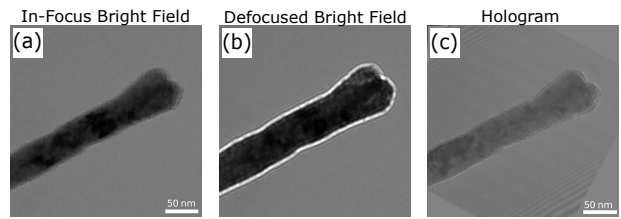

Figure 3: **LTEM bright field micrographs.** Through-focal-series in-focus image (a) and under-focus (b). The scalebar in (a) extends to (b). Off-axis hologram (c) with scalebar of 50 nm.

**2.2.2 Lorentz-mode Scanning Transmission Electron Microscopy** The experimental parameters are presented in Section 5. Figure 4 is a series of images highlighting the acquisition and reconstruction of electron ptychography where the object, the NiFe nanowire, is depicted in a high-angle annular dark field micrograph Fig. 4 (a) at an estimated defocus value of 11 $\mu$m. The convolution of the complex-valued probe, as shown in Fig. 4 (b), with the complex-valued object over an illumined region results in an approximated diffraction intensity, as shown in Fig. 4 (c). The diffraction intensity approximation yields quantitative information to the localized spatial frequencies of the nanowire that contribute to the total electron phase shift. Both electron ptychography and tilt-corrected BF STEM reconstructions were performed utilizing the open-source python 4D-STEM analysis software package, py4DSTEM [32, 36].

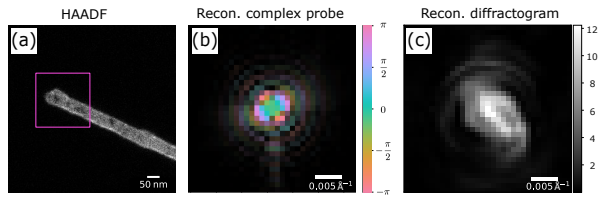

Figure 4: **Ltz-4D-STEM single-slice ptychographic reconstruction**. High-angle annular dark field (HAADF) image (a) reconstructed probe (b), and reconstructed diffractogram (c).

**2.2.3 Comparison of Phase Retrieval Methods**
Finally, we compare the successful reconstructions of the total phase shift using our different approaches: OAH reconstruction as the experimental ground truth basis, RMAD, ePIE, and the simulation as shown in Figure 5. We lay out the remainder of this section as follows: firstly, qualitative (a-d) and quantitative (i) observations of the total phase shift and secondly, qualitative (e-h) and quantitative (j) observations of the phase shift gradient.

First, we note that the experimental ground truth basis (a) and the simulation (d) show similar results for the direction of magnetization as inferred from the projected magnetic induction. The phase shift induced by the magnetization of the nanowire, across its diameter, changes by 2-3 radians which indicates an angular difference of 172°. We find that ePIE is in near-identical agreement with the experimental ground truth basis and simulation, only the direction of the magnetization across the nanowire is reversed. We attribute this inadvertent change in magnetization due to experimental handling of the specimen during insertion and/or removal from one microscope (Ltz-4D-STEM) to the other (LTEM). RMAD, on the contrary, appears vastly different with the appearance of uniform contrast surrounding the length of the nanowire with dark contrast localized to the tip of the nanowire. Additionally, the appearance of a high intensity region on the nanowire extends the total phase shift range by 8 radians. We attribute this shift to a higher sensitivity to strong contrast arising from electron scattering as highlighted by the bright and dark field images in Supplementary Figure 1. We find in the quantitative analysis that the total phase shift range for experimental phase reconstructions using OAH, RMAD, and ePIE are 7.645, 9.442, and 8.219 radians, respectively, whereas the simulated phase reconstruction is only 5.320 radians. We note that this variation in the range of the total phase shift from experimental to simulation can be attributed to a variety of factors including oxidation, carbon build-up, and variations in geometry that may contribute to an increase in the mean inner potential.

Second, we qualitatively analyze the total phase shift gradient which depicts the stray field emanating from the tip of the nanowire 5 (e-h). Starting with a comparison betwen OAH and the simulation represented in Fig. 5 (e and h), we find that the stray fields vary in signal intensity distribution, and, in some cases, orientation. An irregular phase gradient is evident on the right-hand-side of the nanowire in (e) such that it deviates from the ideal simulated case (h) which may be attributed to nearby nanowires, sample charging, or non-uniformity within or around the nanowire. Electron ptychography, on the other hand, depicts a phase gradient in the closest agreement with the simulation of the nanowire, as evidenced by intensity distribution and orientation of the stray fields. Finally, RMAD does not appear to represent a stray field that is physically realistic, evidenced by an upward trajectory of the stray fields away from the nanowire alongside significant oscillations of the phase gradient sign on either sides of the nanowire. Subsequently, we analyzed the difference in the phase shift surrounding the nanowire, which can be directly correlated to the projected magnetic induction of the sample, as highlighted by the region of the phase shift outside the boundaries of the nanowire. We found that OAH and ePIE measures a difference of approximately 2.268 and 2.096 radians, whereas RMAD measures higher at 3.565 radians. The simulation measures a difference of 1.990 radians indicating that quantitatively, OAH - the experimental ground truth basis and ePIE are quantitatively in greater agreement with the simulation, measuring a magnetic phase shift difference of 0.278 and 0.106 radians, respectively.

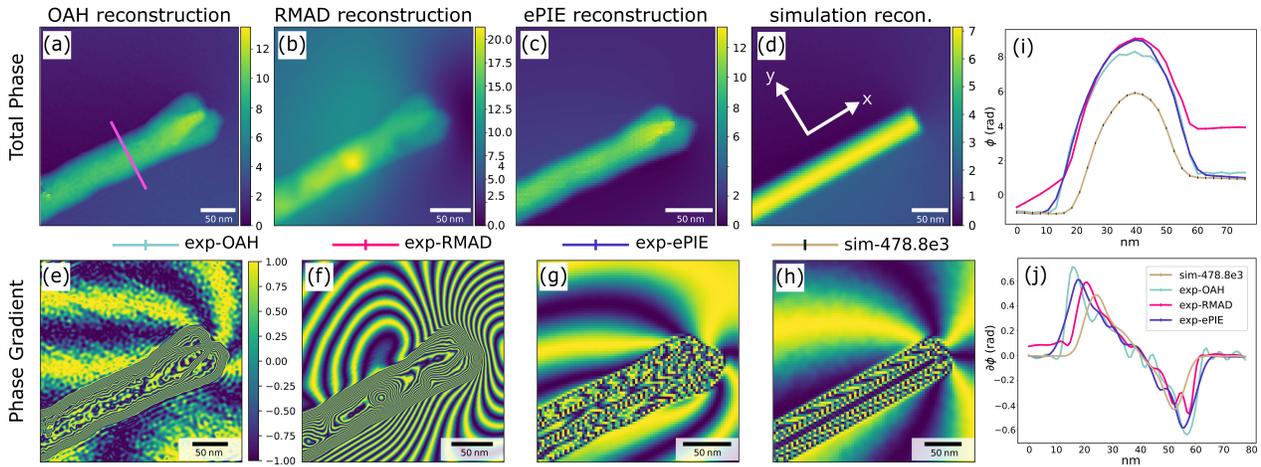

Figure 5: **Comparative performance of the reconstructed total phase shift.** Phase reconstructions for (a) OAH, (b) RMAD, (c) ePIE, and (d) simulation. Gradient of the total phase for (e) OAH, (f) RMAD, (g) ePIE, and (h) simulation. Line profiles are depicted from the position of the line in (a) representing the shift in total phase (i) and the gradient of the total phase (j). Each scalebar is 50 nm.

## 3 Discussion

In this work, we have presented a comparison between phase retrieval for magnetic nanostructures using TIE, RMAD, and ePIE. Based on the simulations of nanoscale islands, the TIE reconstruction showed a loss in spatial resolution and phase accuracy at high defocus valueS. Thus, we rely on OAH to generate an experimental phase shift that we use as our ground truth basis for comparison, and later compare the integrity of automatic differentiation and electron ptychography phase reconstructions against this experimental basis. Simulated phase reconstructions indicate RMAD and ePIE [29] to be comparable in structural similarity and computational resources. However, in assessing the total phase, as presented in Figure 2, electron ptychography is quantitatively more robust than RMAD and maintains high spatial resolution alongside high phase accuracy.

Next, in experimental observations of a NiFe nanowire, we discern three significant findings. First of note, the saturation magnetization of the 36 nm diameter NiFe nanowire was found to be $478.8 \times 10^3$ Amperes/meter, approximately 57% that of Permalloy ($840 \times 10^3$ Amperes/meter) which we used to simulate our conditions initially. Secondly, OAH and ePIE do not vary significantly in reconstructing the total phase shift. However, we note the phase gradient that is representative of the projected magnetic induction is impacted by noise, moreso in OAH than in ePIE, which becomes apparent when comparing to the simulation. Third, RMAD in the experimental application of phase reconstruction did not perform well in retrieving the magnetic contributions of the shift in total phase as depicted in Figure 5 (i). While the total phase shift is in near-agreement with the initial simulation, reflecting on its precision to extracting the mean inner potential with significant sensitivity to geometrical variations, the magnetic contributions ascertained from the phase shift outside the boundaries of the nanowire deviates significantly in qualitative and quantitative observations. We attribute this phase deviation to the algorithm's limitation to strongly scattered electrons arising from geometric irregularities and/or inhomogeneity. Supplementary figure 1 depicts a bright and dark field LTEM micrographs of the NiFe nanowire which highlight these variations.

Additional RMAD investigations using a selection of defocus values are presented in the supplementary materials. Supplementary figure 2 depicts the total phase shift (top row) and the phase gradient (bottom row) from RMAD phase reconstructions using a series of various defocused images within the algorithm. The phase retrieval for the full through-focal-series image stack is depicted in SFIG. 2(a) and highlights phase wrapping on the left-hand-side of the nanowire alongside some slight pixelated regions around the tip. Phase retrieval in RMAD utilizing only the maximally defocused images is presented in SFIG. 2(b); here, phase wrapping and an incomplete reconstruction is evidenced. Subsequently, a combination of the moderate and minimally defocused images are utilized for the phase reconstruction presented in SFIG. 2(c). Here, contrast uniformity along the length of the nanowire is evident with variations occurring on either side which is indicative of a complete reconstruction. A set of minimally defocused images from the dataset were utilized for the reconstruction presented in SFIG.

2(d) which highlights two high intensity signals near the tip of the nanowire with opposing signs, as evidenced by the bright yellow and dark blue spots.

Phase-retrieval using py4DSTEM's parallax algorithm was also performed to evaluate how tilt-corrected bright field STEM compares to electron ptychography as shown in Supplementary Figure 3. The traditional ePIE algorithm phase reconstruction (which we highlight is the limit of stochastic gradient descent for a batch-size=1) is also presented in SFIG. 3(a), highlighting the total phase shift and phase gradient, respectively. SFIG. 3(b) depicts the phase reconstruction using py4DSTEM's stochastic gradient descent ptychography using a batch-size larger than 1, while SFIG. 3(c) depicts the phase reconstruction using py4DSTEM's parallax algorithm, highlighting the total phase shift and phase gradient, respectively. As expected, iterative ptychography using the stochastic gradient descent with a larger batch size yields a reconstruction of the total phase shift that is comparable to ePIE, albeit at smaller computational cost. Yet, in analyzing the phase gradient alongside a quantitative line profile extraction, the results vary from expectation. For instance, the projected magnetic induction to the left of the nanowire appears to deviate in orientation, resolution, and intensity. Additionally, the full-width half maximum (FWHM) of the plotted curve in SFIG. 3(d) is approximately 30 nm, as opposed to ePIE which presents a FWHM of approximately 36 nm which is the consistent with the diameter of the NiFe nanowire. The tilt-corrected bright field STEM reconstruction, on the other hand, reconstructs a total phase shift that is an order of magnitude lower than ePIE and appears to completely suppress the magnetic contributions, as evidenced by the phase gradient in SFIG. 3 (c) and the line extraction in (d). We attribute this to the fact that tilt-corrected BF STEM, unlike ptychography, relies on the weak-phase approximation which is violated due to the strong electrostatic scattering of the NiFe nanowire.

It is apparent that the extended ptychographic iterative engine (ePIE), i.e. stochastic gradient descent with a small batch size, proves to be the most robust in retrieving the object's phase with the highest spatial resolution and phase accuracy, especially as it pertains to magnetic contributions. Experimental investigations and subsequent analyses, however, may be limited by access to Ltz-4D-STEM capabilities and high-performance computing resources necessary to process large 4D-datasets. OAH, on the other hand, is fairly straightforward in computation; however, we note that experimental limitations alongside greater noise and distortion exist. RMAD, while robust in homogeneous films such as those depicted in simulation and experiment [25, 41], its performance in in-homogenous and geometrically constrained materials such as the NiFe nanowire presented in this paper is lacking in spatial resolution and phase accuracy. We assert careful considerations must be taken when performing phase retrieval analysis using electron microscopy for magnetic phase retrieval since it is highly dependent on the instrument, the mean inner potential of the sample, the phase retrieval technique being used, and access to computational resources.

## 4 Conclusions

Quantitative magnetic contributions, from both simulated and experimental conditions, are achievable to extract from total phase shift reconstructions using Lorentz electron microscopy, whether operating in scanning transmission or transmission alone. TIE, while less sensitive to spatial resolution and phase accuracy, remains a solid phase reconstruction method for quick analyses relating to a materials' magnetic induction for both simulation and experiment. Off-axis holography is quantitatively a greater alternative to TIE and RMAD for analysis of the total phase and the magnetic contributions. RMAD, while more straightforward than electron ptychography and computationally less expensive, proved to be the most difficult to attain a fully resolved reconstruction and exhibited the lowest confidence among the experimentally observed phase reconstruction methods. Electron ptychography using ePIE, however, is the phase reconstruction method that provides the greatest confidence in both spatial resolution and phase accuracy in analyzing the total phase and its magnetic contributions.

## 5 Methods

The formalism of TIE is mathematically represented by Equation 2,

$$\nabla_\perp \cdot [I(\mathbf{r}_\perp, 0)\nabla_\perp \varphi(\mathbf{r}_\perp)] = -\frac{2\pi}{\lambda}\frac{\partial I(\mathbf{r}_\perp, z)}{\partial z}\bigg|_{z=0}, \quad (2)$$

where $I(\mathbf{r}_\perp, z)$ is the image intensity at a defocus plane $z$, $\mathbf{r}_\perp$ is a two-dimension (2D) position vector in the image plane, $\lambda$ is the electron wavelength, and $\nabla_\perp$ is the 2D gradient operator.

Section 2.1 utilizes the following parameters for micromagnetic simulations encompassing a region of 826 x 826 x 20 nm$^3$ from a total field-of-view comprised of 5.04 x 5.04 $\mu$m$^2$ x 20 nm. The cell size used was 3.5 x 3.5 x 1 nm. The magnetization of the nanostructures was initially computed using a saturation magnetization of 840×10$^3$ A/m, an exchange stiffness coefficient ($A_{ex}$) of 13e-12 J/m [42], and a Gilbert damping constant ($\alpha$) of 0.02. LTEM images of the simulated magnetization distributions were generated using PyLorentz [18] with microscope parameters as follows: accelerating voltage E = 200 kV, mean inner potential of Permalloy [25] $V_0$ = 26 V, island thickness, t = 20 nm, defocus $\Delta z$ = [$\pm$100, $\pm$90, $\pm$70, $\pm$50, $\pm$30, $\pm$10, 0] $\mu$m, beam coherence $\theta_c$=10 $\mu$rad, and defocus spread of 80 nm [41]. Ltz-4D-STEM dataset was generated using ePIE with microscope parameters equivalent to LTEM simulation parameters presented above. Additional simulation parameters for electron ptychography reconstruction over 306$\times$306 pixels in real- and reciprocal-space, defocus = 100 $mu$m, a probe step size of 25% the calculated probe diameter, and aperture = 500 nm.

Section 2.2.1, relating to the LTEM experiments, utilizes an aberration-corrected JEOL JEM-2100 Lorentz TEM instrument operating at 200 kV to acquire two datasets for OAH and RMAD phase retrieval. The experimental holograms (4096 x 4096 pixels) were acquired at a magnification of 48k X with an applied voltage on the biprism of 70 V, yielding a sampling of 0.137 $\frac{nm}{px}$. The beam stigmators were used to extend the beam perpendicular to the electrical biprism to maximize the coherence along the length of the nanowire. The experimental through-focal-series dataset (4096 x 4096 pixels) needed for RMAD was acquired at 12k X with an exposure of 3 s, yielding a sampling of 0.566 $\frac{nm}{px}$.

Additionally, the simulation corresponding to these experimental observations are established using a nanowire set in a 4.25 x 1.02 x 1.02 $\mu m^3$ region with a cell size of 3.4 x 3.4 x 3.4 nm. The magnetization of the nanowire was calculated using parameters consistent with Permalloy as described previously.

Section 2.2.2, relating to the Ltz-4D-STEM experiment, was performed using a JEOL NEOARM cold-FEG STEM instrument operating at 200 kV in Lorentz-mode (low-mag mode). The acquired 4D-STEM dataset used for ptychographic reconstruction was collected with PNDetector's pnCCD (S)TEM camera and has 256$\times$256 probe positions with 0.0009 $\mu$m step size. The recorded diffraction patterns have 264$\times$264 pixels in size and a sampling of 0.0258 $\frac{mrad}{px}$. Defocus of the dataset is estimated to be 11 $\mu$m. The data analysis corresponding to Section 2.2.2 utilizes py4DSTEM's single-slice ptychography reconstruction with a batch-size=1 which is the traditional ePIE algorithm. We note the following parameters used for the reconstructions: reciprocal-binning = 8, rolloff = 0.015, iterations = 100000, step size = 0.25, and the experimentally derived semi-angle = 0.237 mrad. While these are presented here, we note that these will vary based on individual datasets. The binning parameter was initiated to reduce the dataset and subsequently, reduce computation expense. A rolloff was initiated to construct a soft aperture to avoid oscillatory behavior, known as Gibbs ringing, due to sharp edges. A fixed step size of 0.25 was chosen to ensure slow convergence to the minimum.

The extended analysis presented in Section 3 for SFIG. 3 utilized py4DSTEM's parallax algorithm for tilt-corrected BF-STEM. The technique utilizes the principle of reciprocity between virtual STEM images formed from off-axis pixels in the BF-disk to tilted plane-wave illumination. Briefly, the algorithm proceeds as follows: i) creates a stack of virtual BF images using pixels in the BF-disk; ii) cross-correlate and shift the virtual BF images to each other to obtain the in-plane shifts arising due to the defocused tilted illumination; iii) fit the measured in-plane shifts to the gradient of the aberration surface to estimate low-order aberrations (namely defocus); iv) phase-flip the negative-valued parts of the BF contrast transfer function (CTF), to ensure the final aligned stack has dark-field like contrast. Unlike ptychography, parallax reconstructions are less sensitive to hyperparameters, with the py4DSTEM default values working reasonably well in this case.

## 6 Acknowledgments

This work was funded by the U.S. Department of Energy, Office of Science, Office of Basic Energy Sciences, Materials Science and Engineering Division. H. N. and M. C. were supported by a DOE-BES Early Career project FWP No. ERKCZ55 (H.N.). Lorentz 4D-STEM was performed at the ORNL's Center for Nanophase Materials Sciences (CNMS), which is a DOE Office of Science User Facility. We gratefully acknowledge the computing resources provided on Swing, a high-performance computing cluster operated by the Laboratory Computing Resource Center at Argonne National Laboratory.

## 7 Data Availability

The raw data acquired for this study is publicly available in a Zenodo repository at DOI: 10.5281/zenodo.14503519. The code used to process the raw datasets in this study will be shared in a public repository on GitHub.

## 8 Competing Interests

The authors declare that they have no competing interests.

## 9 Copyright

# 10 SUPPLEMENTARY FIGURES

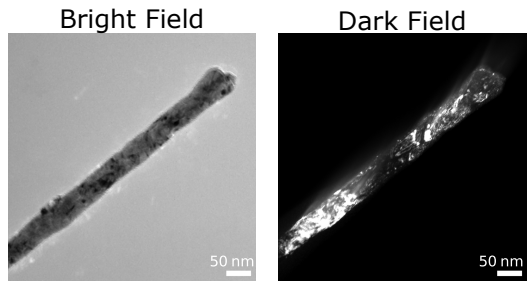

**SFIG. 1: LTEM micrographs.** Bright field **(left)** and dark field **(right)** of the NiFe nanowire used for phase retrieval. Scalebar is 50 nm.

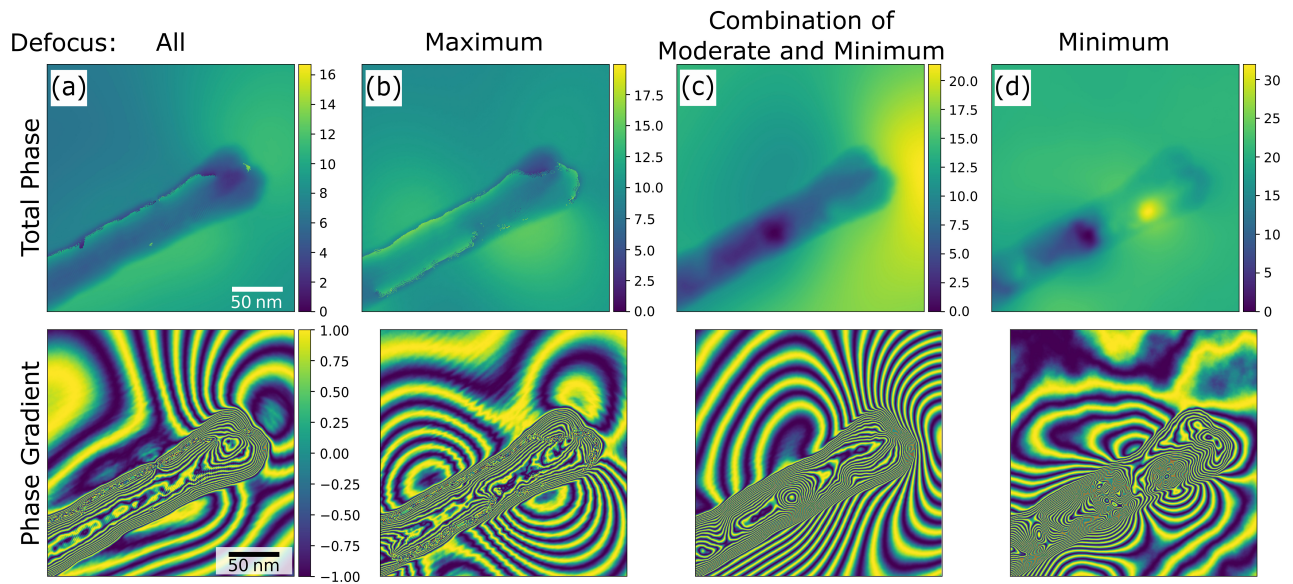

**SFIG. 2: RMAD phase shift reconstructions using ADLTEM software.** Total phase shift (top row) and total phase shift gradient (bottom row) are reconstructed using a series of defocused conditions from the experimentally acquired through-focal-series dataset containing (a) all, (b) maximum, (c) combination of moderate and minimum, and (d) minimum defocused images. Scalebar in top row (white) and bottom row (black) are 50 nm and applicable to each image within the row.

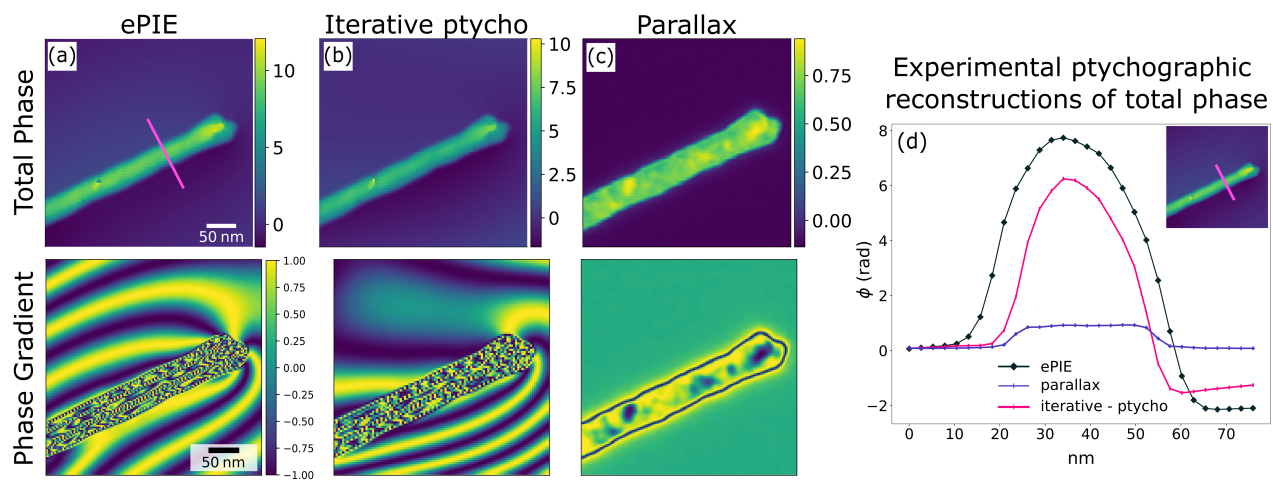

**SFIG. 3: 4D STEM phase reconstructions.** Reconstructed total phase shifts (top row) and phase gradients (bottom row) using ePIE (a), iterative ptychography with GDO (b), and parallax (c). Line profile of the total phase (d) from the line position highlighted in (a). Scalebars are 50 nm and applicable to images within the row.